# A Neural Network Approach for the Peak Profile Characterization

Ruben A. Dilanian

diruben@optusnet.com.au

The neural network–based approach, presented in this paper, was developed for the analysis of peak profiles and for the prediction of base profile characteristics, such as width, asymmetry, asymptotes ("peak tales"), *etc.* of the observed distributions. The obtained parameters can be used as the initial parameters in the peak decomposition applications. The neural network architecture, presented here, was designed for the analysis of one particular type of peak profiles, the Voigt-type distributions (symmetrical and asymmetrical), and is suitable for a variety of applications, such as x-ray and neutron powder diffraction, x-ray spectroscopy, *etc*. The approach itself, however, is not limited to the demonstrated case, but is applicable to other types of peak profile distributions. The approach was successfully tested on experimentally collected x-ray powder diffraction data.

## 1. Introduction

Peak profile analysis is employed in many research areas, such as physics, chemistry, material science, bioinformatics, *etc*. The main purpose of the analysis is the extraction of important information from the observed distributions. The accuracy of the analysis significantly affects the outcomes of the corresponding research. The core idea of any peak profile analysis approach (pattern fitting, peak profile decomposition) is to describe the observed distribution by a well-defined analytical function or by a combination of such functions. The set of adjustable parameters of the peak profile function defines the main characteristics of the observed distribution. The more complex the profile of the observed peak, the more complicated the combination of analytical functions used, leading to more adjustable parameters being incorporated into the analysis. Consequently, the peak profile characterization process becomes more complex.

The critical point of the peak profile characterization, that makes the analysis particularly difficult, is the asymmetry of the observed distributions, as well as the behaviors of peak "tails". The latter, having a weak magnitude and broad distribution, carry important information, the loss or incorrect treatment of which leads to significant misinterpretation of the obtained data. This subject has already been discussed many times (Rietveld, 1967; Rao & Gopalakrishan, 1997; Felinger, 1998; Gubicza, 2014) and, more importantly, a variety of peak profile fitting/decomposition applications have been developed. In the course of this, various types of peak profile functions and their combinations have been proposed, tested and incorporated into the analysis.

Despite of the large number of peak profile analysis applications, the success of the analysis depends on several crucial, particularly for the inexperienced user, questions: what kind of peak profile function is most suitable for this particular analysis? How many adjustable parameters are necessary for the profile characterization and how does one choose their initial values?

These questions fit well into a more general classification problem, when the decision is made on the basis of certain rules and criteria, which, in turn, are based on previously accumulated information (experience). This is the area in which the neural network applications have shown their effectiveness and robustness (see, for example, Hagan *et al.*, 1996; Rojas, 1996; Kriesel, 2005 and references therein). The neural network, with the proven capability to classify objects or tasks according to predefined rules, can be used for the analysis and characterizations of measured peak distributions. The main goal of the current research is to develop a neural network−based approach for the peak profile characterization and analysis, a network that can accumulate crucial information throughout the learning process and works on a base of predefined rules and criteria.

In this article, I look at one particular side of the proposed approach: prediction of the base profile characteristics, such as width, asymmetry, asymptotes ("peak tales"), *etc.* of the observed distributions. To be more specific, only the Voigt-type (symmetrical and asymmetrical) peak profiles are considered in this case. It should be noted, however, that the proposed approach is applicable to other types of distribution.

For now, let us consider an observation consisted of a set of peaks with the Voigt-type peak profiles. The role of the neural network application is to define the main characteristics of the observed peaks, which can further be used as an initial guess in the peak decomposition applications.

## 2. Case study: the Voigt-type peak profile

Consider a distribution, $y_m(x)$, which is isolated from other measurements within the predefined region, $x \in \mathbf{R}$, and consists of a set of $N$ discrete values $\{y_m(x_1), y_m(x_2), \ldots, y_m(x_N)\}$. The given distribution can be approximated by the pseudo−Voigt function, which is a linear combination of the Lorentzian and the Gaussian functions

$$V(x|\mu, H, \eta) = \eta L(x|\mu, H) + (1-\eta)G(x|\mu, H), \qquad (1a)$$

$$L(x|\mu, H) = \frac{1}{\pi}\frac{2}{H}\left(1 + 4\left(\frac{x-\mu}{H}\right)^2\right)^{-1}, \qquad (1b)$$

$$G(x|\mu,H) = \frac{1}{\sqrt{\pi}} \frac{\sqrt{4\ln(2)}}{H} \exp\left(-4\ln(2)\left(\frac{x-\mu}{H}\right)^2\right), \qquad (1c)$$

where $\mu$ is the location parameter (or the peak position), $H$ is the full-width at the half-maximum (FWHM) and $\eta$ is the fraction parameter. Although Eqs. (1) are suitable for the symmetrical distributions only, it is not critical for the following analysis. Later in the text I will shown how to extend the outcomes of the analysis to the asymmetrical profile case. Our task is to define a set of profile parameters, $\{\mu_0, H_0\}$, of the two given functions, $G(x|\mu,H)$ and $L(x|\mu,H)$, as well as the $\eta_0$ parameter, which satisfy the condition

$$y(x) = y_m(x) - y_B(x) = y_o\{\eta_0 L(x|\mu_0,H_0) + (1-\eta_0)G(x|\mu_0,H_0)\}, \qquad (2a)$$

or

$$y(x) = y_o V(x|\mu_0, H_0, \eta_0). \qquad (2b)$$

Here, $y_o$ is the amplitude of the measured distribution and $y_B(x)$ is the background. To simplify things, let us assume for now that the background is a constant, $y_B(x) = y_B$, for all $x \in \mathbf{R}$ and the location parameter, $\mu$, is known (the corresponding index will be omitted from the following equations).

The given functions, $G(x|\mu,H)$, $L(x|\mu,H)$ and $V(x|\mu_0,H_0,\eta_0)$, as well as the target function, $y(x)$, can be considered as vectors, $\mathbf{g}_H$, $\mathbf{l}_H$, $\mathbf{v}_H$ and $\mathbf{y}$, in the $N$–dimensional vector space. The corresponding angles between these vectors can be defined as:

$$\cos(\varphi_{GY}) = \frac{\mathbf{g}_H^T \mathbf{y}}{|\mathbf{g}_H||\mathbf{y}|}, \quad \cos(\varphi_{LY}) = \frac{\mathbf{l}_H^T \mathbf{y}}{|\mathbf{l}_H||\mathbf{y}|}, \qquad (3a)$$

$$\cos(\varphi_{GL}) = \frac{\mathbf{g}_H^T \mathbf{l}_H}{|\mathbf{g}_H||\mathbf{l}_H|}, \quad \cos(\varphi_{VY}) = \frac{\mathbf{v}_{H,\eta}^T \mathbf{y}}{|\mathbf{v}_{H,\eta}||\mathbf{y}|}, \qquad (3b)$$

where $|\mathbf{g}_H|^2 = \mathbf{g}_H^T\mathbf{g}_H = d_G^2$, $|\mathbf{l}_H|^2 = \mathbf{l}_H^T\mathbf{l}_H = d_L^2$, $|\mathbf{v}_{H,\eta}|^2 = \mathbf{v}_{H,\eta}^T\mathbf{v} = d_V^2$, $|\mathbf{y}|^2 = \mathbf{y}^T\mathbf{y} = d_Y^2$, and $T$ denotes the transpose operation. The alignments of three vectors, $\mathbf{g}_H$, $\mathbf{l}_H$ and $\mathbf{v}_H$, are governed by the values of the corresponding peak profile parameters, while the alignment of the $\mathbf{y} = \mathbf{y}_m - \mathbf{y}_B$ vector depends only on accuracy of the measurement and the background. According to Eq. (2b), two vectors, $\mathbf{y}$ and $\mathbf{v}_{H,\eta}$, will be aligned parallel to each other, $\varphi_{VY} = 0$, when $H = H_0$ and $\eta = \eta_0$. In this case, Eq. (2b) can be rewritten in the following vector form:

$$\mathbf{y} = y_0 \mathbf{v}_{H_0,\eta_0} = y_0\left(\eta_0 \mathbf{l}_{H_0} + (1-\eta_0)\mathbf{g}_{H_0}\right). \qquad (4)$$

Moreover, Eq. (3a) can also be modified by replacing the vector **y** with the vector $\mathbf{v}_{H,\eta}$ (note that this replacement is correct only when $H = H_0$ and $\eta = \eta_0$):

$$\cos(\varphi_{\mathbf{GV}}) = \frac{\sum_{x \in \mathbf{X}} g_{x,H}[\eta l_{x,H} + (1-\eta)g_{x,H}]}{d_G d_V} = \eta \frac{d_L}{d_V}\cos(\varphi_{\mathbf{GL}}) + (1-\eta)\frac{d_G}{d_V} \quad (5a)$$

$$\cos(\varphi_{\mathbf{LV}}) = \frac{\sum_{x \in \mathbf{X}} l_{x,H}[\eta l_{x,H} + (1-\eta)g_{x,H}]}{d_L d_V} = \eta \frac{d_L}{d_V} + (1-\eta)\frac{d_G}{d_V}\cos(\varphi_{\mathbf{GL}}). \quad (5b)$$

From the sum of these two equations it follows that:

$$\cos(\varphi_{\mathbf{GV}}) + \cos(\varphi_{\mathbf{GL}}) = (1 + \cos(\varphi_{\mathbf{GL}}))\left[\eta \frac{d_\mathbf{L}}{d_\mathbf{V}} + (1-\eta)\frac{d_\mathbf{G}}{d_\mathbf{V}}\right] \quad (6a)$$

or

$$\cos\left(\frac{\varphi_{\mathbf{GV}} + \varphi_{\mathbf{LV}}}{2}\right)\cos\left(\frac{\varphi_{\mathbf{GV}} - \varphi_{\mathbf{LV}}}{2}\right) = \left(\cos\left(\frac{\varphi_{\mathbf{GL}}}{2}\right)\right)^2 \left[\eta \frac{d_\mathbf{L}}{d_\mathbf{V}} + (1-\eta)\frac{d_\mathbf{G}}{d_\mathbf{V}}\right]. \quad (6b)$$

This leads to the following two conditions, which define the relationship of the three vectors, $\mathbf{g}_H$, $\mathbf{l}_H$ and **y**, in the case when $H = H_0$ and $\eta = \eta_0$:

$$\begin{cases} \cos\left(\dfrac{\varphi_{\mathbf{GV}} + \varphi_{\mathbf{LV}}}{2}\right) = \cos\left(\dfrac{\varphi_{\mathbf{GL}}}{2}\right) \\ \cos\left(\dfrac{\varphi_{\mathbf{GV}} - \varphi_{\mathbf{LV}}}{2}\right) = A\cos\left(\dfrac{\varphi_{\mathbf{GL}}}{2}\right) \end{cases}, \quad (7)$$

where $A = (\eta d_\mathbf{L} + (1-\eta)d_\mathbf{G})/d_\mathbf{V}$. From the first condition, which can also be rewritten as $\varphi_{\mathbf{GV}} + \varphi_{\mathbf{LV}} = \varphi_{\mathbf{GL}}$, it follows that for a given vector, $\mathbf{v}_{H_0}$, the Gaussian, $\mathbf{g}_H$, and the Lorentzian, $\mathbf{l}_H$, vectors will vary with $H$ in such a way that $\varphi_{\mathbf{GV}} + \varphi_{\mathbf{LV}} \to \varphi_{\mathbf{GL}}$, when $H \to H_0$. When the first condition is satisfied and the FWHM parameter, $H_0$, is estimated, the $\eta_0$ parameter is defined by the second condition:

$$\eta_0 = \left(-d_\mathbf{G} + d_\mathbf{V}\frac{\cos((\varphi_{\mathbf{GV}} - \varphi_{\mathbf{LV}})/2)}{\cos((\varphi_{\mathbf{GV}} + \varphi_{\mathbf{LV}})/2)}\right)(d_\mathbf{L} - d_\mathbf{G})^{-1}. \quad (8)$$

So far, only the symmetrical profile functions have been considered. To introduce asymmetry into the analysis, the methodology of the split pseudo–Voigt function (Toraya, 1990) can be applied. The two independent sets of profile parameters, $(\eta, H)$, are assigned to the left, $x < \mu$, and the right, $x > \mu$, sides of the $y_m(x)$ distribution. In this case, the conditions, Eqs. (7), are applied separately to both sides of the measured peak profile. The ratio of the two estimated $H$ defines the asymmetry of the measured reflections, $A = H_0(x<\mu)/H_0(x>\mu)$.

It should be noted that Eq. (3) defines the angles between the normalized vectors $\mathbf{g}_H/|\mathbf{g}_H|$, $\mathbf{l}_H/|\mathbf{l}_H|$, and $\mathbf{y}/|\mathbf{y}|$. Consequently, the Gaussian and Lorentzian functions can be replaced by their non-normalized analogues:

$$L(x) = \left(1 + \left(r_L B|x-\mu|\right)^2\right)^{-1} \tag{9a}$$

$$G(x) = \exp\left(-\left(r_G B|x-\mu|\right)^2\right), \tag{9b}$$

where $B = H^{-1}$, $r_L = 2$, $r_G = \sqrt{4\ln(2)}$ and $x \in \mathbf{X}$. Eqs. (9 and 7) form the foundation of the neural network architecture, which will be described in the next section.

## 3. Neural network based peak profile analyser (NNPA).

In this section, I will follow the notations accepted in (Hagan *et al.*, 1996). The complete NNPA network is shown in Fig. 1. The network consists of four layers: the Input layer, the Output layer and two hidden layers, Layer 1 and Layer 2, respectively.

Layer 1 is a radial basis network [Powell, 1987]. The net input for Layer 1 is the $N$–dimensional vector:

$$\mathbf{n}^{(1)} = \left|\mathbf{p} - W^{(1)}\right| b^{(1)}, \tag{10}$$

where $\mathbf{p} = \mathbf{X}$ is an $N$–dimensional input vector, $W^{(1)} = \mu$, and $b^{(1)} = B = H^{-1}$ are the weight and bias of Layer 1, respectively. The superscript in brackets indicates the number of the layer. The two transfer functions, $G(\mathbf{n})$ and $L(\mathbf{n})$, are activate the response of Layer 1:

$$\mathbf{a}^{(1)} = (\mathbf{a}_G, \mathbf{a}_L)^T, \tag{11a}$$

where

$$\mathbf{a}_G = \exp\left(-\left(r_G \mathbf{n}^{(1)}\right)^2\right), \tag{11b}$$

$$\mathbf{a}_L = (1 + \left(r_L \mathbf{n}^{(1)}\right)^2)^{-1}. \tag{11c}$$

Then, the responses of the transfer functions, shown in Fig. 2, are normalized to produce the following output of Layer 1:

$$\mathbf{A}_R = (\mathbf{A}_G, \mathbf{A}_L)^T, \tag{12}$$

where $\mathbf{A}_G = \mathbf{a}_G/|\mathbf{a}_G|$ and $\mathbf{A}_L = \mathbf{a}_L/|\mathbf{a}_L|$.

The second hidden layer, Layer 2, is a network with two neurons, depicted by two subscript indices, 0 and 1, two $\mathrm{acos}(\mathbf{n})$ transfer functions and zero biases. The weight of the first neuron, $\mathbf{W}_1^{(2)}$, is a 2 by $N$ matrix, each row of which is equal to the corresponding row of the output of Layer 1, i.e. $\mathbf{W}_1^{(2)} = \mathbf{A}_R$.

The net input of this neuron is an inner product between the weights, $\mathbf{W}_1^{(2)}$, and the normalized input vector $\mathbf{p}_1^{(2)} = \mathbf{Y}/|\mathbf{Y}| = \mathbf{y}_m/d_Y$:

$$\mathbf{n}_1^{(2)} = \left(\sum_{j=1}^{N} W_{1,i,j}^{(2)} p_j^{(2)}\right)^T \quad for\ i = 1, 2, \tag{13a}$$

or

$$\mathbf{n}_1^{(2)} = \left(\mathbf{A}_G^T \mathbf{p}_1^{(2)}, \mathbf{A}_L^T \mathbf{p}_1^{(2)}\right)^T. \tag{13b}$$

The response of the first neuron is:

$$a_1^{(2)} = \varphi_{GY} + \varphi_{LY}, \tag{14}$$

where $\varphi_{GY} = \mathrm{acos}(\mathbf{A}_G^T \mathbf{p}_1^{(2)})$ and $\varphi_{LY} = \mathrm{acos}(\mathbf{A}_L^T \mathbf{p}_1^{(2)})$. The weight of the second neuron, $\mathbf{W}_0^{(2)}$, is equal to $\mathbf{A}_G$ and the corresponding input, $\mathbf{p}_0^{(2)}$, is equal to $\mathbf{A}_L$. The bias of this neuron also equals zero. Consequently, the net input of this neuron is an inner product between the two outputs of Layer 1:

$$n_0^{(2)} = \sum_{i=1}^{N} W_{0i}^{(2)} p_{0,i}^{(2)} = \mathbf{A}_G^T \mathbf{A}_L \tag{15}$$

and the corresponding response is

$$a_0^{(2)} = \varphi_{GL} = \mathrm{acos}(n_0^{(2)}). \tag{16}$$

Figure 3 shows the response of Layer 2 as a function of the $H$. The symmetrical target profile, $y(x)$, was simulated within the range of $\mu_0 - 2.0 \leq x \leq \mu_0 + 2.0$ with the following parameters: $\mu_0 = 15.0$, $H_0 = 0.1$ and $\eta_0 = 0.5$, $y_0 = 650$ and $y_B = 320$. The FWHM was varied in the range: $0.02 \leq H \leq 0.2$. As one can see from the figure, the angle between $\mathbf{g}_H$ and $\mathbf{l}_H$, $\varphi_{GL}(H)$, is a slowly varying function, while $\varphi_{GY}(H) + \varphi_{LY}(H)$ exhibits the strong minima when $H = H_0$ and when the condition $\varphi_{GY} + \varphi_{LY} = \varphi_{GL}$ is satisfied.

A close examination of Layer 2 response near the critical point, $H_0$, reveals, however, that there is a region surrounding $H_0$ where the values of the two responses of Layer 2, $a_1^{(2)}$ and $a_0^{(2)}$, are extremely close to each other, which limits the accuracy of the critical point determination. This is due to the fact that NNPA is designed to implement only the $\varphi_{GY} + \varphi_{LY} = \varphi_{GL}$ condition (see Eq. (7)). A detailed analysis of the behavior of the NNPA shows that the uncertainty in the determination of the critical point using only one condition does not exceed 5 percent of the nominal value.

The Output layer, shown in Fig. 1 as a box with the letter C, is a standard single neuron linear network, Fig. 4a, which defines the difference between the two outputs of Layer 2. The net input of this neuron is $n^{(O)} = a_1^{(2)} - a_0^{(2)}$ and the corresponding response is $a^{(O)} = n^{(O)}$, or

$$A_C = (\varphi_{GY} + \varphi_{LY}) - \varphi_{GL}. \tag{17}$$

According to the condition, $\varphi_{GY} + \varphi_{LY} = \varphi_{GL}$, the target output of the NNPA network, Eq. (17), should be zero, $t_{NNPA} = 0$. In the case of the asymmetric peak profile, the NNPA network is applied separately to both sides of the profile, $x < \mu$ and $x > \mu$ for all $x \in \mathbf{X}$. It is much more convenient, therefore, to analyze the response of the NNPA network by the following parameter:

$$F(0,0) == (A_C(x < \mu), A_C(x > \mu)), \tag{18}$$

where the double zeros indicate the target values of the net output for both sides of the peak profile.

There are two supervised training approaches that can be used with NNPA. The first is the backpropagation approach (see Hagan *et al.*, 1996, for more details), when the initial value of $B_{ini}$ is iteratively modified according to the following rule (the steepest descent algorithm):

$$B_{k+1} = B_k - \alpha \frac{\partial E}{\partial B}, \tag{19}$$

until the error cost function, $E = |t_{NNPA} - A_C|^2 = |a_2^{(2)} - a_0^{(2)}|^2$, is minimized. In Eq. (19), $\alpha$ is the learning rate and $k$ is the iteration number. Since the output $a_2^{(2)}$ of Layer 2 exhibits only one strong minimum in a wide range of $H$, the convergence of this algorithm is guarantied [Hagan *et al.*, 1996]. The second approach involves searching for the minimum output of the NNPA network, $\min\{A_C(H)\}$, from a set of input values of $H$, generated within the predefined range $H_{min} < H < H_{max}$. Both approaches yield similar results in a course of the training process. However, the second approach is more appropriate for the profile parameters prediction, since it does not require the derivative calculations.

As was previously mentioned, NNPA uses the normalized responses from Layer 1, $\mathbf{A}_G$ and $\mathbf{A}_L$, as well as the normalized input vector $\mathbf{Y}$ to produce the output, Eq. (17). The normalization operator, shown as a box with the letter N in Fig. 1, was used to achieve this. The normalization operator, Fig. 4b, is a single neuron network with the inverse square root transfer function, $f(n) = n^{-1/2}$. The net input of the network is an inner product between the weights, $\mathbf{W}^{(N)}$, and the input vector, $\mathbf{p}$,

$$n^{(N)} = (\mathbf{W}^T \mathbf{p}) = \sum_{x \in \mathbf{X}} W_x p_x = \sum_{x \in \mathbf{X}} p_x^2, \quad (20a)$$

since $\mathbf{W}^{(N)} = \mathbf{p}$. Consequently, the response of this network is

$$\mathbf{a}^{(N)} = \mathbf{p}/\sqrt{n^{(N)}} = \mathbf{p}/|\mathbf{p}|. \quad (20b)$$

## 4. Applications.

In this section, the performance of NNPA is demonstrated. First, NNPA was tested on two simulated asymmetrical peak profiles. Then, NNPA was used to analyse several measured x-ray powder diffraction patterns, two of which (synchrotron data of $BaSO_4$ and $C_{10}H_{16}N_6S$) are from the templates of the RIETAN-FP Rietveld refinement program (Izumi & Momma, 2007), and the other (synchrotron data of the Malaria Parasite Pigment, Hemozoin) is from (Klonis *at al.*, 2010). It should be noted that NNPA can be applied to any kind of distribution and is not limited to x-ray powder diffraction data. The analysis consisted of three steps. First, the NNPA network was applied to estimate the peak profile parameters, $H$ (x < μ), $H$ (x > μ), and η of the selected reflections. Then, the obtained parameters were used to calculate the peak profiles, $y_c(x)$, for all the selected reflections (see Eqs. (1, 2)). The background and the amplitude were estimated in the following way:

$$y_B = \min\{y_m(x)\} - \beta \min\{V(x|\mu, H, \eta)\}, \quad (21a)$$

$$y_0 = \frac{\int_{x \in \mathbf{X}} [y_m(x) - y_B] dx}{\int_{x \in \mathbf{X}} [V(x|\mu, H, \eta)] dx}, \quad (21b)$$

where $\beta > 1$ is the correction factor.

Finally, the calculated profile was compared with the original one, $y_m(x)$. The quality of the analysis was controlled by the following *R*-factor:

$$R = \frac{1}{Z} \sum_{x \in \mathbf{X}} |y_m(x) - y_c(x)| \quad (22a)$$

and the corresponding error cost function

$$E_y = \frac{1}{Z} \sum_{x \in \mathbf{X}} (y_m(x) - y_c(x))^2, \quad (22b)$$

where $Z = \sum_{x \in \mathbf{X}} y_m(x)$.

*Simulated data*

Two asymmetrical peak profiles where simulated using the parameters shown in Table 1. The NNPA gives the following estimated parameters (*R*-factor = 0.01):

(Sample 1) $H(x < \mu) = 0.121$, $H(x > \mu) = 0.030$, $\eta = 0.67$ and $F(0,0) = (0.0, 0.0)$;

(Sample 2) $H(x < \mu) = 0.041$, $H(x > \mu) = 0.011$, $\eta = 0.31$ and $F(0,0) = (0.0, 0.0)$.

Based on the obtained parameters, the peak profiles, $y_c(x)$, were generated and compared with the simulated data, $y_m(x)$, Fig 5.

*Experimental data*

The results of the NNPA analysis of the experimental data are collected in Table 2. The corresponding peak profiles are shown in Figs. 6, 7 and 8, which are in agreement with measured data.

Some notes about the obtained results should be added. Eq. (21a) assumes the constant background under the peak, which, in general, is not realistic. Since NNPA analyses the peak profile of individual reflections within the local area around the peak position, the background can be accurately represented as a linear function $y_B(x) = y_B^0 + y_B^1(x)$. As one can see, only the constant part of the background function, $y_B^0$, has been used in the analysis so far, while its linear part, $y_B^1(x)$, has been ignored. This approach leads to the relatively high values of the $F(0,0)$ parameter (see, for example, results for $C_{10}H_{16}N_6S$). Moreover, the differences between the two values of $F(0,0)$ indicate the behaviour of the background around the selected reflection.

To fix this problem, two additional parameters, $\Delta_1^B$ and $\Delta_2^B$, where introduced into Layer 2 of NNPA. The net input of the neuron (1) is modified in the following way:

$$\mathbf{n}_1^{(2)} = \left(\sum_{j=1}^{N} W_{1,i,j}^{(2)} \tilde{p}_j^{(2)}\right)^T \quad for\ i = 1, 2, \tag{23}$$

where $\tilde{p}_j^{(2)} = p_j^{(2)} + \Delta_1^B$ for $x < \mu$ and $\tilde{p}_j^{(2)} = p_j^{(2)} + \Delta_2^B$ for $x > \mu$. In this way, the background is adjusted separately for both sides of the peak profile. Two parameters are iteratively modified according to the following rules:

$$\begin{aligned}\Delta_{1,k+1}^B &= \Delta_{1,k}^B + \delta \frac{A_C(x < \mu)}{|A_C(x < \mu)|} \\ \Delta_{2,k+1}^B &= \Delta_{2,k}^B + \delta \frac{A_C(x > \mu)}{|A_C(x > \mu)|}\end{aligned} \tag{24}$$

until the error cost function is minimized, $E = |t_{NNPA} - A_C| < E_{\min}$. Here, $\delta$ is the learning rate and $k$ is the iteration number. The effect of the background adjustment using the Eq. 23 is demonstrated for $C_{10}H_{16}N_6S$ in Table 3 (compare with Table 2).

Another reason for the relatively high values of the $F(0,0)$ parameter can be the incorrect estimation of the peak position, $\mu$, which can also be adjusted, if necessary, in a similar way as the background was. Thereby, the response of NNPA can be adjusted by two factors, the background correction and the peak position correction, see Fig. 8.

## 5. Conclusion

The neural network approach for the preliminary characterisation of peak profiles has been developed and tested on measured distributions. The main assumption of the method is that the observed distribution can be approximated by two analytical functions, the Gaussian and the Lorentzian. The first one describes the central part of the peak distribution, while the second one defines the asymptotes ("tales") of the observed peak. The core idea of the approach is to represent the measured distribution, as well as the two analytical functions as vectors in the $N$-dimensional vector space, where $N$ is the number of measured points. The relative alignment of these vectors is determined by the peak profile parameters of the two analytical functions. The specific alignment conditions, derived in Section 2, define the correct values of the peak profile parameters. Despite the fact that NNPA was designed for the analysis of Voigt-type profiles, it is applicable to a broad range of measured distributions, such as x-ray and neutron powder diffraction, x-ray spectroscopy, *etc*. Moreover, the proposed approach is suitable for the analysis of other types of peak profiles, which will be shown in further publications.

The NNPA java application is available on request (diruben@optuanet.com.au).

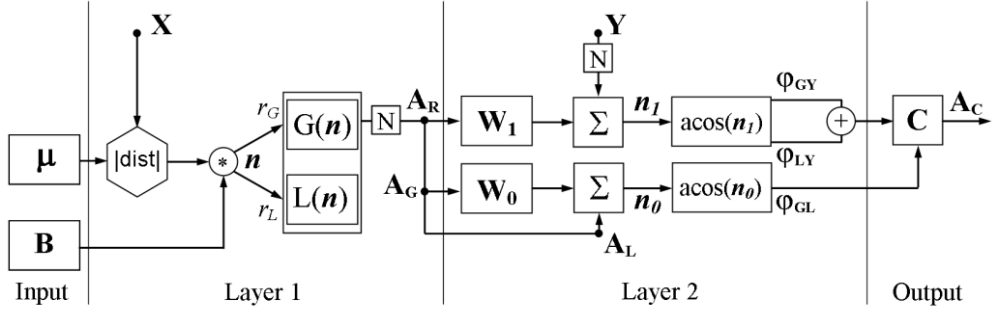

**Figure 1**. The complete structure of the NNPA network.

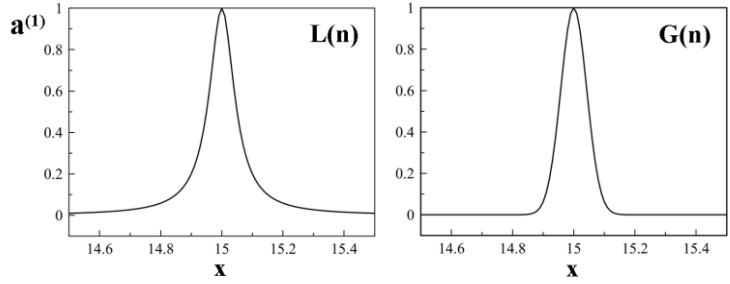

**Figure 2**. The Lorentzian, **L**(**n**), and the Gaussian, **G**(**n**), response of Layer 1

($H = 0.1$, $\mu = 15.0$).

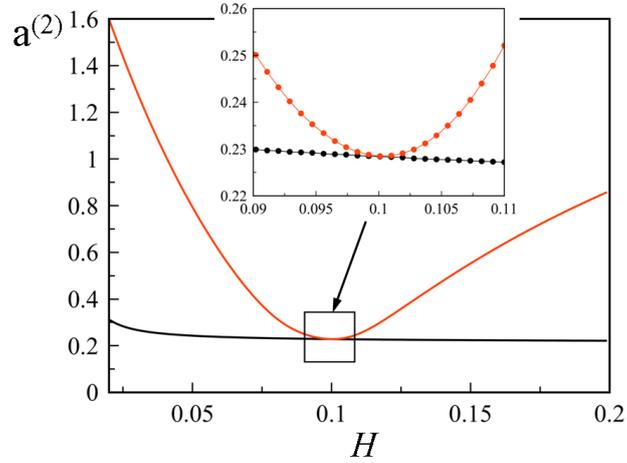

**Figure 3**. The response of Layer 2 as a function of FWHM;

(black line): $\varphi_{GL}(H)$, (red line): $\varphi_{GY}(H) + \varphi_{LY}(H)$.

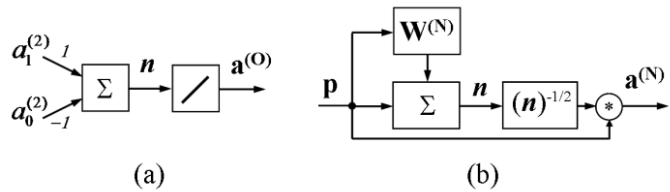

**Figure 4**. (a) The Output layer of the NNPA network. (b) The normalization operator.

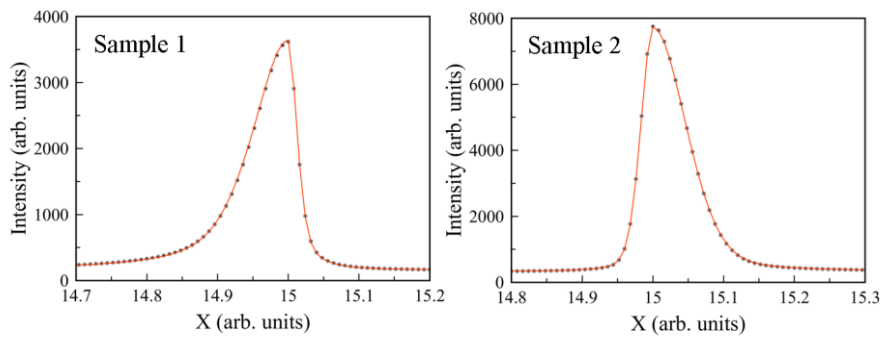

**Figure 5**. The peak profile simulated (black markers) and generated (red lines) using the NNPA approach.

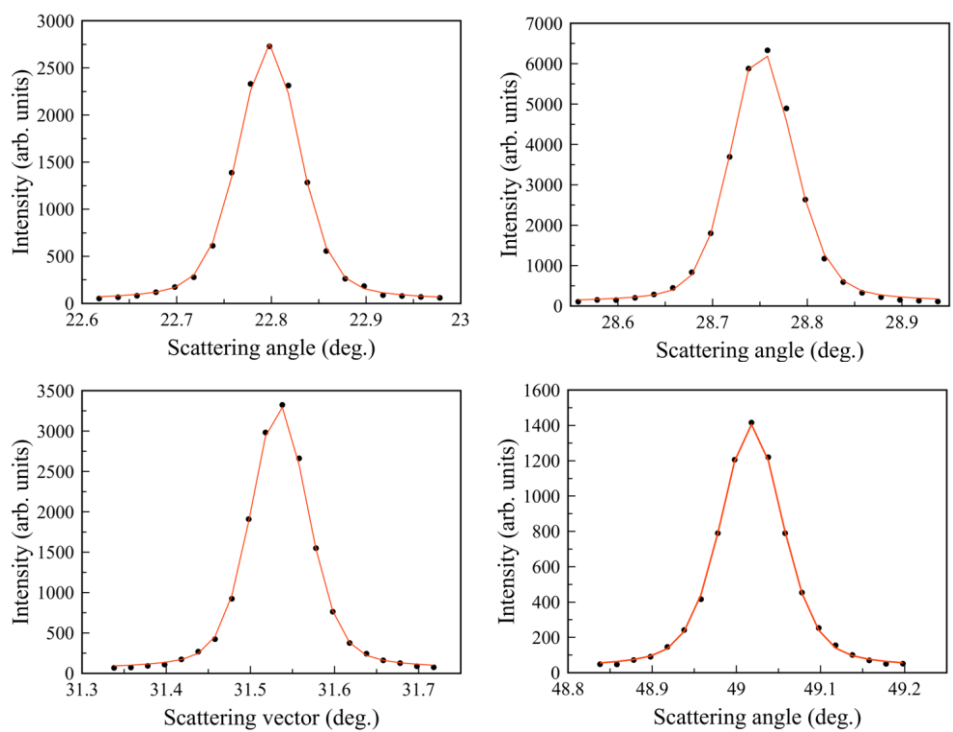

**Figure 6**. Selected reflections from the x-ray powder diffraction pattern of BaSO$_4$; measured data (black markers), the peak profile generated using NNPA (red line).

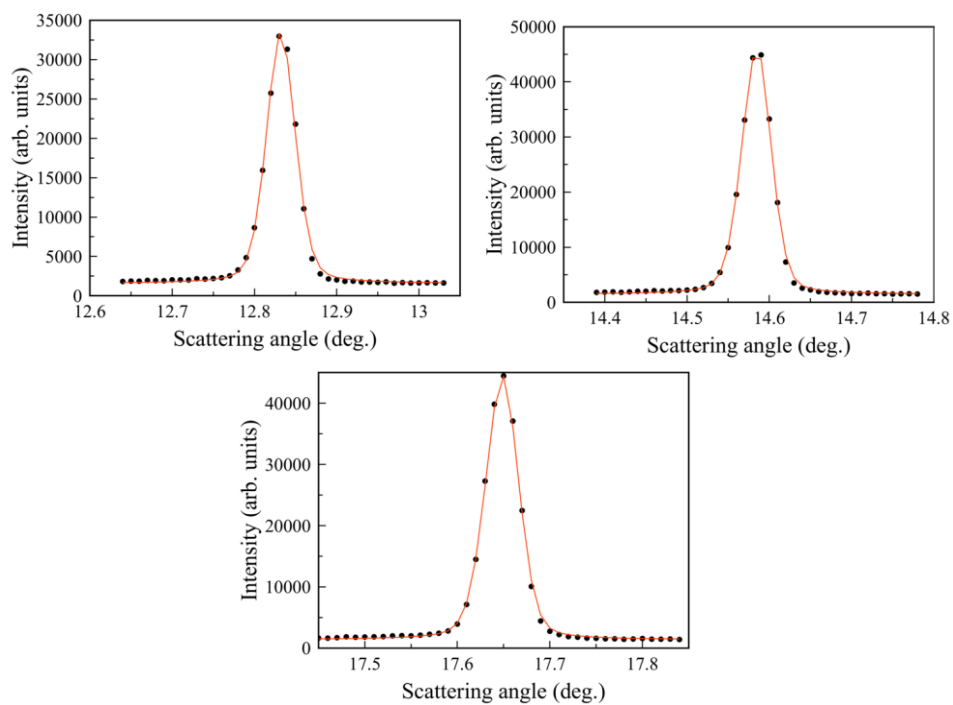

**Figure 7**. Selected reflections from the x-ray powder diffraction pattern of $C_{10}H_{16}N_6S$; measured data (black markers), the peak profile generated using NNPA (red line).

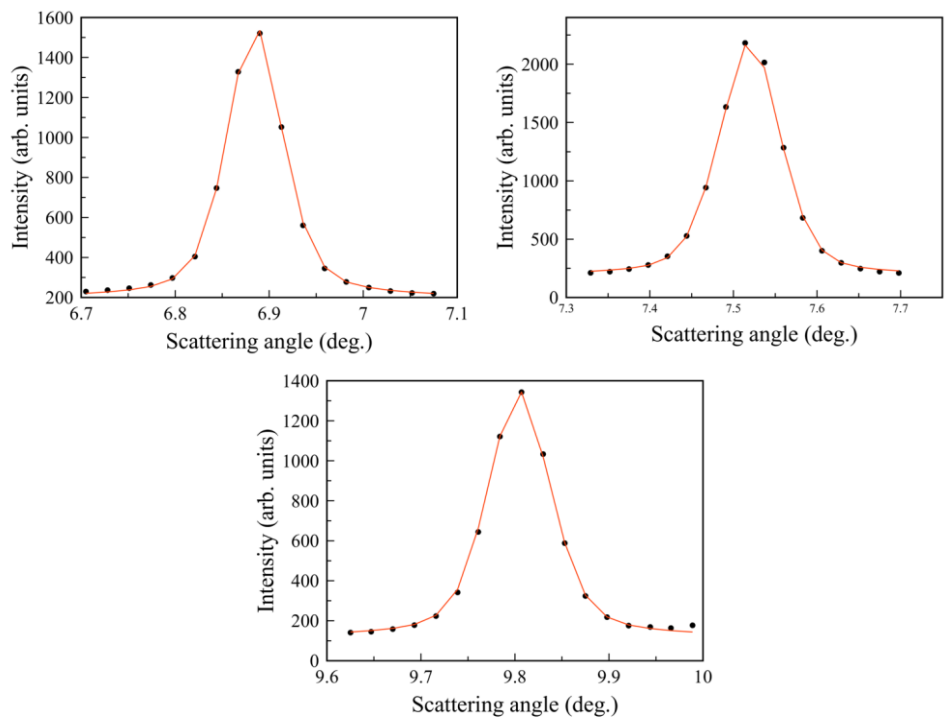

**Figure 8**. Selected reflections from the x-ray powder diffraction pattern of Hemozoin; measured data (black markers), the peak profile generated using NNPA (red line).

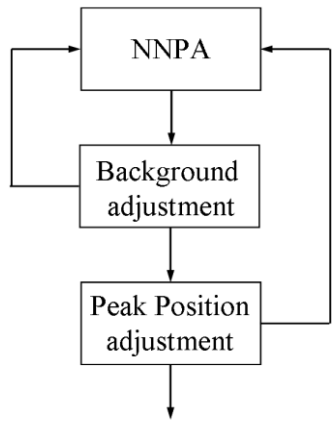

**Figure 9**. Schematic representation of the complete NNPA network.

Table 1. Peak profile parameters for simulated data.

| Sample | X range | $\mu_0$ | $H_0(x<\mu)$ | $H_0(x>\mu)$ | $\eta$ | $y_0$ | $y_B$ |
|---|---|---|---|---|---|---|---|
| 1 | 14 ÷ 16 | 15 | 0.12 | 0.03 | 0.7 | 357 | 155 |
| 2 | 14 ÷ 16 | 15 | 0.04 | 0.11 | 0.3 | 657 | 321 |

Table 2. Estimated peak profile parameters

| **BaSO$_4$** | | | | |
|---|---|---|---|---|
| $\mu$ | 22.798 | 28.750 | 31.534 | 49.018 |
| $H(x<\mu)$ | 0.078 | 0.073 | 0.079 | 0.088 |
| $H(x>\mu)$ | 0.075 | 0.083 | 0.082 | 0.088 |
| $<H>$ | 0.077 | 0.078 | 0.081 | 0.088 |
| $\eta$ | 0.43 | 0.40 | 0.43 | 0.50 |
| $F(0,0)$ | (0.003, 0.008) | (0.002, 0.008) | (0.004, 0.002) | (0.004, 0.002) |
| $R$-factor | 0.038 | 0.040 | 0.030 | 0.024 |
| **C$_{10}$H$_{16}$N$_6$S** | | | | |
| $\mu$ | 12.832 | 14.585 | 17.649 | |
| $H(x<\mu)$ | 0.041 | 0.043 | 0.043 | |
| $H(x>\mu)$ | 0.042 | 0.041 | 0.041 | |
| $<H>$ | 0.042 | 0.042 | 0.042 | |
| $\eta$ | 0.36 | 0.33 | 0.32 | |
| $F(0,0)$ | (0.002, 0.012) | (0.002, 0.009) | (0.004, 0.005) | |
| $R$-factor | 0.053 | 0.038 | 0.040 | |
| **Hemozoin** | | | | |
| $\mu$ | 6.884 | 7.520 | 9.804 | |
| $H(x<\mu)$ | 0.069 | 0.087 | 0.078 | |
| $H(x>\mu)$ | 0.072 | 0.085 | 0.080 | |
| $<H>$ | 0.070 | 0.086 | 0.079 | |
| $\eta$ | 0.56 | 0.45 | 0.54 | |
| $F(0,0)$ | (0.001, 0.001) | (0.001, 0.002) | (0.002, 0.006) | |
| $R$-factor | 0.012 | 0.018 | 0.019 | |

**Table 3**. Estimated peak profile parameters of $C_{10}H_{16}N_6S$ after the background adjustment ($\delta = 0.02$, $E_{min} = 5e^{-4}$).

| | | | |
|---|---|---|---|
| μ | 12.832 | 14.585 | 17.649 |
| $H(x < \mu)$ | 0.041 | 0.043 | 0.043 |
| $H(x > \mu)$ | 0.042 | 0.041 | 0.041 |
| $<H>$ | 0.042 | 0.042 | 0.042 |
| η | 0.36 | 0.33 | 0.32 |
| $F(0,0)$ | (0.000, 0.004) | (0.000, 0.000) | (0.000, 0.000) |
| $R$-factor | 0.053 | 0.038 | 0.040 |